\documentclass[conference,10pt]{IEEEtran}
\IEEEoverridecommandlockouts
\usepackage{amsmath,amsthm,amssymb,color,cite,graphicx}
\usepackage{algorithm}
\usepackage{algorithmic}
\usepackage{subfigure}
\usepackage{bm}
\usepackage{balance}
\usepackage{booktabs}
\usepackage{fancyhdr} % for adding the copyright notice

\makeatletter
\newcommand*{\rom}[1]{\expandafter\@slowromancap\romannumeral #1@}
\makeatother

\def\thetabf{\boldsymbol \theta}

\def\ebf{{\bf e}}

\def\hbf{{\bf h}}

\def\nbf{{\bf n}}

\def\sbf{{\bf s}}

\def\ubf{{\bf u}}

\def\wbf{{\bf w}}

\def\Ibf{{\bf I}}

\def\Bc{{\cal B}}

\def\Kc{{\cal K}}

\def\Mc{{\cal M}}

\def\Pc{{\cal P}}

\def\Sc{{\cal S}}

\def\ie{{\it i.e.,\ \/}}
\def\nn{\nonumber}

\def\Re{\mathfrak{R}\mathfrak{e}}
\def\Im{\mathfrak{I}\mathfrak{m}}

\def\mae{{\mathbb{E}}}

\theoremstyle{definition}

\newtheorem{assumption}{Assumption}
\newtheorem{proposition}{Proposition}

\begin{document}

\title{Multi-Model Wireless Federated Learning with Downlink Beamforming}

\author{ Chong Zhang$^{\star}$, Min Dong$^{\dagger}$, Ben Liang$^{\star}$, Ali Afana$^{\ddagger}$, Yahia Ahmed$^{\ddagger}$\\
\normalsize $^{\star}$Dept.\ of Electrical and Computer Engineering, University of Toronto, Canada, $^{\ddagger}$Ericsson Canada, Canada   \\
$^{\dagger}$Dept.\ of Electrical, Computer and Software Engineering, Ontario Tech University, Canada
}%

\maketitle

%\IEEEpubidadjcol

\thispagestyle{fancy}

\fancyhead[L]{}
\fancyhead[C]{}
\fancyhead[R]{}
\lfoot{}
\cfoot{\scriptsize \copyright 2023 IEEE. Personal use of this material is permitted. Permission from IEEE must be obtained for all other uses, in any current or future media, including reprinting/republishing this material for advertising or promotional purposes, creating new collective works, for resale or redistribution to servers or lists, or reuse of any copyrighted component of this work in other works.}
\renewcommand{\headrulewidth}{0mm}

\begin{abstract}
This paper studies the  design of wireless federated learning (FL) for simultaneously training multiple machine learning models.
We consider round robin device-model assignment and downlink  beamforming for concurrent  multiple model updates. After formulating the joint downlink-uplink transmission process,
we derive the per-model global update expression over communication rounds, capturing the effect of beamforming and noisy reception.
To maximize the multi-model\ training convergence rate, we derive an upper bound on the optimality gap of the global model update and use it to formulate a multi-group multicast beamforming problem. We show that this problem can be converted to minimizing the sum
of inverse received signal-to-interference-plus-noise ratios, which can be solved efficiently by projected gradient descent.
Simulation  shows that our proposed multi-model
FL solution outperforms other alternatives, including conventional single-model sequential training and multi-model zero-forcing beamforming.
\end{abstract}

\section{Introduction}
\label{sec:intro}
Federated learning (FL) \cite{Mcmahan&etal:2017} is a widely adopted method for multiple devices to collaboratively train a common machine learning (ML)\ model.
In wireless FL, a parameter server, usually the base station (BS), uses wireless communication to exchange model parameters with participating devices \cite{Zhu&etal:2020}.
With the frequent exchange of a large number of parameters, FL performance degrades in the wireless environment due to
signal distortion and limited wireless resources.
This necessitates efficient communication design to effectively support FL.

Most existing works on wireless FL have focused on training only a single model \cite{Chen&etal:JSAC2012FLWN,Yang&etal:TWC2020,Zhu&etal:TWC2020,Zhang&Tao:TWC2021,Sun&etal:JSAC2022,Fan&etal:TWC2022,Amiri&etal:TWC2022,Wei&Shen,Guo&etal:JSAC2022,Wang&etal:JSAC2022b,Zhang&etal:arxiv2023}.
Assuming an error-free downlink, \cite{Chen&etal:JSAC2012FLWN,Yang&etal:TWC2020,Zhu&etal:TWC2020,Zhang&Tao:TWC2021,Sun&etal:JSAC2022,Fan&etal:TWC2022}
focused on efficient transmission for the uplink acquisition of local parameters from devices to the BS, including both digital \cite{Chen&etal:JSAC2012FLWN}
and analog \cite{Yang&etal:TWC2020,Zhu&etal:TWC2020,Zhang&Tao:TWC2021,Sun&etal:JSAC2022,Fan&etal:TWC2022} schemes.
Noisy downlink transmission for FL was studied in \cite{Amiri&etal:TWC2022} with error-free uplink.
It was shown in these works that analog transmission can be more efficient than digital for both the downlink and the uplink.
Joint noisy downlink-uplink transmission for FL was studied in  \cite{Wei&Shen,Guo&etal:JSAC2022,Wang&etal:JSAC2022b,Zhang&etal:arxiv2023},
with single-antenna BSs in \cite{Wei&Shen,Guo&etal:JSAC2022,Wang&etal:JSAC2022b}
and a multi-antenna BS in \cite{Zhang&etal:arxiv2023}.
In practice, the parameter server may have multiple models to be trained.
Directly using the existing single-model FL schemes  may lead to substantial latency,
degrading the overall performance of wireless FL.

Simultaneously training multiple models in FL was proposed recently in \cite{Bhuyan&etal:2023},
assuming error-free downlink and uplink transmissions.
It is shown that multi-model FL can substantially improve the training convergence rate over the single-model FL approach,
which reduces the burden of required computation and communication.
However, the idealized system in \cite{Bhuyan&etal:2023} did not account for
the impact of wireless transmission over noisy channels.
In multi-model wireless FL, besides noisy downlink and uplink transmission during model updates,
there is also inter-model interference in transmission, which adds substantial challenges to improving  communication efficiency.

This paper studies multi-model wireless FL design for noisy downlink and uplink wireless channels
with a multi-antenna BS.
We consider analog transmission, downlink beamforming, and round robin model scheduling.
Aiming to maximize the training convergence rate, we derive an upper bound on the optimality gap of the FL global model
update, which captures the impact of noisy transmission and inter-model interference.
We then show that the minimization of this upper bound leads to a downlink multi-group multicast beamforming design to minimize the sum of inverse received signal-to-interference-plus-noise ratios (SINRs)
subject to a downlink transmit power budget at the BS, which can be solved using projected gradient descent (PGD).
Our simulation results under typical wireless network settings show that the proposed multi-model FL design substantially outperforms the conventional  approach
of sequentially training one model at a time and the multi-model training using the popular
zero-forcing beamforming scheme.

\allowdisplaybreaks
\section{System Model}\label{sec:system_prob}
\subsection{Multi-Model FL System}\label{sec:fl_model}
We consider FL in a wireless network consisting of a server and $K$
worker devices that collaboratively train $M$ global models at the server.
Let  $\Kc_{\text{tot}} = \{1, \ldots, K\}$ denote the total set of devices and  $\Mc = \{1, \ldots, M\}$  the set of models.
Let $\thetabf_m\in\mathbb{R}^{D_m}$ be the parameter vector of model $m$ with $D_m$ parameters,
and assume $\|\thetabf_m\| < \infty$.
Assume each device $k \in \Kc_{\text{tot}}$ holds local training datasets for all $M$ models,
with each $\thetabf_m$ being locally trained using the dataset for model $m$ of size $S_m^k$, denoted by $\Sc_m^{k} = \{(\sbf_{m,i}^k,v_{m,i}^k): 1 \le i \le S_m^k\}$, where
$\sbf_{m,i}^k\in\mathbb{R}^{b} $ is the $i$-th data feature vector and $v_{m,i}^k$ is the label for this data sample.
The local training loss function representing the training error
at device $k$ for model $m \in \Mc$ is defined as
$F_m^{k}(\thetabf_m)=\frac{1}{S_m^k}\sum_{i=1}^{S_m^k} L_m(\thetabf_m;\sbf_{m,i}^k,v_{m,i}^k)$,
where $L_m(\cdot)$ is the sample-wise training loss for model $m$.
 The global training loss function for model $m$ is given by the weighted sum of the local loss functions for model $m$ over all $K$ devices:
$F_m(\thetabf_m) = \frac{1}{\sum_{k=1}^{K}S_m^{k}}\sum_{k=1}^{K}S_m^{k}F_m^{k}(\thetabf_m)$.
The learning objective is to find optimal $\thetabf_m^\star$ that minimizes $F_m(\thetabf_m)$ for
each model $m\in\Mc$.

\begin{figure}[t]
\centering
\hspace{-1em}\includegraphics[scale=.43]{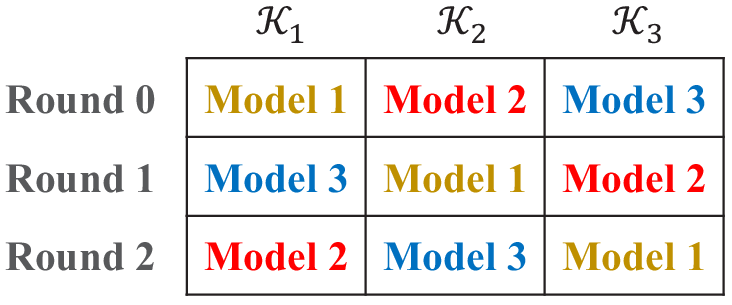}
\caption{An example of round robin scheduling of device-model assignment in a frame for training 3 models.}\vspace*{-1.8em}
\label{fig1:rr}
\end{figure}

The $K$ devices use their respective local training datasets to simultaneously train the $M$ models
and communicate with the server via noisy downlink and uplink wireless channels to exchange the model training information iteratively.
At the beginning of each downlink-uplink communication round $t=0,1,\ldots$, the devices are divided into device groups, and the server assigns the training task of each model to a device group. We consider the round robin scheduling approach for efficient device-model assignment \cite{Bhuyan&etal:2023}. Specifically, we define every $M$ communication rounds as a \emph{frame}. At the beginning of each frame, the $K$ devices are partitioned    randomly into $M$ equal-sized groups. Let $\Kc_{i}$ denote the set of devices in device group $i=1,\ldots,M$. These device groups remain unchanged within a frame.
Each  device group $i$ is assigned to train  model $\hat{m}(i, t)$ at round $t$ given by
\begin{align}\label{m(i)}
\hat{m}(i, t) = [(M + i - [t \,\,\text{mod}\,\, M] - 1) \,\,\text{mod}\,\, M] + 1.
\end{align}
Fig.~\ref{fig1:rr} shows an example of the device-model assignment in a frame via the round robin scheduling  with $M=3$
models.

The iterative multi-model FL training procedure in each downlink-uplink communication round $t$ is then given as follows:
\begin{itemize}
\item \emph{Downlink broadcast}: The server  broadcasts each of the current $M$ global model parameter vectors $\thetabf_{m,t}$'s to its assigned device group via the downlink channel;
\item  \emph{Local model update}: Device $k\in\Kc_i$ in device group $i$ is scheduled to locally train model $\hat{m}(i, t)$  using its corresponding local dataset $\Sc_{\hat{m}(i, t)}^{k}$.
In particular, the device divides $\Sc_{\hat{m}(i, t)}^{k}$ into mini-batches for its local model update based on $\thetabf_{\hat{m}(i, t),t}$, where it performs $J$ iterative local updates and generates the updated local model $\thetabf^{k,J}_{\hat{m}(i, t),t}$;
\item \emph{Uplink aggregation}: The devices send their updated local models $\thetabf^{k,J}_{m,t}$'s to the server via the uplink  channels.
The server aggregates $\{\thetabf^{k,J}_{m,t}\}_{k\in\Kc_i}$ received from each device group $i$ to generate updated global model $\thetabf_{m,t+1}$, $m \in \Mc$, for the next communication round $t+1$.
\end{itemize}

\subsection{Wireless Communication Model}
We consider a practical wireless communication system where
 the server is hosted by a BS equipped with $N$ antennas, and each device has a single antenna. To efficiently send $M$ global model updates,  the BS uses the multi-group multicast beamforming technique \cite{Dong&Wang:TSP2020,Zhang&etal:TSP2023}
 to send the $M$ global model updates $\thetabf_{m,t}$'s to $M$ device groups  simultaneously over a common downlink channel.
Also, we consider analog transmission, where the BS sends the values of  $\thetabf_{m,t}$'s directly under its transmit power budget.
For the uplink aggregation, we consider the orthogonal multiple access technique to efficiently use the system bandwidth for local model aggregation at the BS. For each device group, we consider over-the-air
computation via analog aggregation over the multiple access
channel. Specifically, the devices in a device group $i$ send their local models  $\{\thetabf^{k,J}_{m,t}\}_{k\in\Kc_i}$   simultaneously over the same uplink channel, while the channels among device groups are orthogonal to each other.

The received model updates over downlink are the distorted
noisy versions of $\thetabf_{m,t}$'s, due to the inter-group interference in transmitting $\thetabf_{m,t}$'s and the noisy communication channel. The uplink received model updates are also the distorted noisy versions of $\thetabf^{k,J}_{m,t}$'s due to the  noisy channel. These errors in the model updates further propagate over subsequent communication rounds for multi-model training, degrading the learning performance. In this paper, we focus on the communication aspect of FL multi-model training
and develop the downlink beamforming design to maximize the learning performance of FL over wireless transmissions.

\section{Multi-Model Downlink-Uplink Transmissions }\label{sec:FL_alg}
In this section, we formulate the wireless transmission process in downlink and uplink for the multi-model update
using the three steps in a communication round that are mentioned in Section~\ref{sec:fl_model}.

\subsection{Downlink Broadcast}\label{subsec:dl_broadcast}
At the start of round $t$, the BS has the  current global models, with model $m$ denoted by
$\thetabf_{m,t} = [\theta_{m1,t}, \ldots, \theta_{mD_m,t}]^T$.
For efficient transmission, we  represent $\thetabf_{m,t}$ using a complex signal vector, whose real and imaginary parts respectively contain the first and second half of the elements in $\thetabf_{m,t}$. That is, $\thetabf_{m,t} = [(\tilde{\thetabf}_{m,t}^{\text{re}})^{T}, (\tilde{\thetabf}_{m,t}^{\text{im}})^T]^T$, where
$\tilde{\thetabf}_{m,t}^{\text{re}} \triangleq [\theta_{m1,t}, \ldots, \theta_{m({D_m}/{2}),t}]^T$ and $\tilde{\thetabf}_{m,t}^{\text{im}} \triangleq [\theta_{m({D_m}/{2}+1),t}, \ldots, \theta_{mD_m,t}]^T$. Let   $\tilde{\thetabf}_{m,t}$ denote the equivalent complex vector representation of $\thetabf_{m,t}$. It is given by
$\tilde{\thetabf}_{m,t} = \tilde{\thetabf}^{\text{re}}_{m,t} + j\tilde{\thetabf}^{\text{im}}_{m,t}\in \mathbb{C}^{\frac{D_m}{2}}$.

With the frame structure,  round $t$ is in frame $n = \lfloor t/M\rfloor$. We assume the downlink channel remains unchanged within one frame. Thus, let $\hbf_{k,n}\in\mathbb{C}^{N}$ be the downlink channel vector from the BS
to device $k \in \Kc_i$, $i=1,\ldots,M$, in frame $n$, which is assumed to be known perfectly at the BS.
Let $\wbf^{\text{dl}}_{i,n}\in\mathbb{C}^{N}$  be  the downlink multicast beamforming vector for device group $i$ in frame $n$.
The BS uses $\wbf^{\text{dl}}_{i,n}$ to  send the normalized complex global model $\frac{\tilde{\thetabf}_{\hat{m}(i,t),t}}{\|\tilde{\thetabf}_{\hat{m}(i,t),t}\|}$ to  device group $i$.
Let $D_{\text{max}}\triangleq\max_{m\in\Mc} D_m$.
 The $M$ model updates are simultaneously sent using $\frac{D_{\text{max}}}{2}$ channel uses. For model $m$ with  $D_{m} < D_{\text{max}}$, the BS randomly sets the  position for $\tilde{\thetabf}_{m,t}$ within $\frac{D_{\text{max}}}{2}$ channel uses and
applies zero padding to the rest of the positions. Thus, the transmitted signal vector for  model $m$ is
$\bar{\thetabf}_{m,t} = [{\bf{0}}^H, \tilde{\thetabf}^H_{m,t}, {\bf{0}}^H]^H$.
 Assume  $\hat{m}(i, t)=m$.
The received signal $\ubf_{k,t}$ at device $k \in \Kc_i$
corresponding to $\tilde{\thetabf}_{m,t}$  is given by
\begin{align}
\ubf_{k,t} \!=\! (\wbf^{\text{dl}}_{i,n}\!)^H\!\hbf_{k,n}\frac{\tilde{\thetabf}_{m,t}}{\|\tilde{\thetabf}_{m,t}\|} \!+\!\! \sum_{j\neq i}(\wbf^{\text{dl}}_{j,n}\!)^H\!\hbf_{k,n}\frac{\bar{\thetabf}^{'}_{\hat{m}(j, t),t}}{\|\tilde{\thetabf}_{\hat{m}(j, t),t}\|} \!+\!\nbf^{\text{dl}}_{k,t}  \nn
\end{align}
where $n = \lfloor t/M\rfloor$,
$\bar{\thetabf}^{'}_{\hat{m}(j, t),t}\in\mathbb{C}^{\frac{D_m}{2}} $ is the portion of  $\bar{\thetabf}_{\hat{m}(j, t),t}$ that aligns with the location of $\tilde{\thetabf}_{m,t}$ in $\bar{\thetabf}_{m,t}$ due to zero-padding, and
$\nbf^{\text{dl}}_{k,t}\sim \mathcal{CN}({\bf 0}, \sigma^2_{\text{d}}\Ibf)$ is  the receiver additive white Gaussian noise (AWGN) vector.
The beamforming vectors $\{\wbf^{\text{dl}}_{i,n}\}_{i=1}^M$ are subject to the BS transmit power budget. Let $D_{\text{max}}P$ be the BS total transmit power budget for sending the entire normalized global models in $D_{\text{max}}$ channel uses, where $P$ denotes the average power
budget per channel use. Then, we have the  transmit power constraint $\sum^{M}_{i=1}\|\wbf^{\text{dl}}_{i,n}\|^2 \leq D_{\text{max}}P$.
 The BS also sends the scaling factor $\frac{\hbf^H_{k,n}\wbf^{\text{dl}}_{i,n}\|\tilde{\thetabf}_{m,t}\|}{|\hbf^H_{k,n}\wbf^{\text{dl}}_{i,n}|^2}$ to  the device via the downlink signaling channel to facilitate this receiver processing.
After post-processing   $\ubf_{k,t}$ using the received scaling factor at device $k\in\Kc_i$, we have
\begin{align}
\!\!& \hat{\tilde{\thetabf}}^{k}_{m,t} = \frac{\hbf^H_{k,n}\wbf^{\text{dl}}_{i,n}\|\tilde{\thetabf}_{m,t}\|}{|\hbf^H_{k,n}\wbf^{\text{dl}}_{i,n}|^2}\ubf_{k,t} \nn\\
\!\!\!\!\!\!&  = \!\tilde{\thetabf}_{m,t} \!+\! \sum_{j\neq i}\!\frac{\hbf^H_{k,n}\wbf^{\text{dl}}_{i,n}(\wbf^{\text{dl}}_{j,n})^H\hbf_{k,n}}{|\hbf^H_{k,n}\wbf^{\text{dl}}_{i,n}|^2}\!\cdot\!
\frac{\|\tilde{\thetabf}_{m,t}\|\bar{\thetabf}^{'}_{\hat{m}(j, t),t}}{\|\tilde{\thetabf}_{\hat{m}(j, t),t}\|}\! + \! \tilde{\nbf}^{\text{dl}}_{k,t}
\label{dl_device_signal}
\end{align}
where $\tilde{\nbf}^{\text{dl}}_{k,t} \triangleq \frac{\hbf^H_{k,n}\wbf^{\text{dl}}_{i,n}\|\tilde{\thetabf}_{m,t}\|}{|\hbf^H_{k,n}\wbf^{\text{dl}}_{i,n}|^2}\nbf^{\text{dl}}_{k,t}$
is the post-processed noise vector at device $k\in \Kc_i$.
By the equivalence of real and complex signal representations   $\thetabf_{m,t}$ and $\tilde{\thetabf}_{m,t}$,
device $k\in \Kc_i$ obtains the estimate of the global model $\thetabf_{m,t}$ as
\begin{align}
\hat{\thetabf}^k_{m,t} & = \big[\Re{\big\{\hat{\tilde{\thetabf}}^k_{m,t}\big\}^T}, \Im{\big\{\hat{\tilde{\thetabf}}^k_{m,t}\big\}^T}\big]^T.
\label{eq_dl}
\end{align}

\subsection{Local Model Update}\label{subsec:device_update}
Device $k \in \Kc_i$ is scheduled to perform local model updates on $\hat{\thetabf}^k_{m,t}$ in \eqref{eq_dl}
using its local dataset $\Sc_{m}^{k}$.
We assume each device adopts the standard mini-batch stochastic gradient descent (SGD) algorithm \cite{Bubeck:2015convex} to perform
the local model training.
In particular,  assume that each device applies $J$ mini-batch SGD iterations for its local model update
in each communication round.
Let $\thetabf^{k,\tau}_{m,t} $ denote the local model update by device $k\in \Kc_i$ at iteration $\tau \in \{0,\ldots,J-1\}$, with
$\thetabf^{k,0}_{m,t} = \hat{\thetabf}^k_{m,t}$ and   $\Bc^{k,\tau}_{m,t}$  the mini-batch  at iteration  $\tau$, which is a subset of  $\Sc^k_{m}$.
The local model update is given by
\begin{align}
\thetabf^{k,\tau+1}_{m,t} & = \thetabf^{k,\tau}_{m,t} - \eta_n \nabla F_{m}^{k}(\thetabf^{k,\tau}_{m,t}; \Bc^{k,\tau}_{m,t}) \nn\\
& = \thetabf^{k,\tau}_{m,t} - \frac{\eta_n}{|\Bc^{k,\tau}_{m,t}|}\sum_{(\sbf,v)\in\Bc^{k,\tau}_{m,t}}\nabla L_{m}(\thetabf^{k,\tau}_{m,t}; \sbf,v) \label{SGD}
\end{align}
where $\eta_n$ is the learning rate in frame $n$,
$\nabla F_{m}^{k}$ and $\nabla L_{m}$ are the gradients of the corresponding loss functions  for model $m$ w.r.t. $\thetabf^{k,\tau}_{m,t}$.
After  $J$ iterations, the device obtains the updated local model $\thetabf^{k,J}_{m,t}$.

\subsection{Uplink Aggregation}\label{subsec:ul_aggre}
The devices send their updated local models $\thetabf^{k,J}_{m,t}$'s to
the BS via the uplink channels.
For efficient transmission,  we again represent  $\thetabf^{k,J}_{m,t}$ using a complex vector, similar to downlink transmission.
That is, we  re-express $\thetabf^{k,J}_{m,t} = [(\tilde{\thetabf}^{k,J\text{re}}_{m,t})^T, \, (\tilde{\thetabf}^{k,J\text{im}}_{m,t})^T]^T$, where
$\tilde{\thetabf}^{k,J\text{re}}_{m,t}$ and  $\tilde{\thetabf}^{k,J\text{im}}_{m,t}$ contain the first and second half of elements in $\thetabf^{k,J}_{m,t}$, respectively.
The equivalent complex vector representation of $\thetabf^{k,J}_{m,t}$ is thus  given by
$\tilde{\thetabf}^{k,J}_{m,t} = \tilde{\thetabf}^{k,J\text{re}}_{m,t} + j\tilde{\thetabf}^{k,J\text{im}}_{m,t}\in\mathbb{C}^{\frac{D_{m}}{2}}$.

We adopt the orthogonal multiple access technique in uplink to efficiently use the system bandwidth for local model aggregation at the BS.
In particular, devices in the same group $i$ send their local model updates $\{\tilde{\thetabf}^{k,J}_{m,t}\}_{k\in\Kc_i}$ to the BS simultaneously via the same uplink channel. The channels among device groups are orthogonal to each other.
Thus, for each model $m$, the BS aggregates the received local model updates from the corresponding assigned device group $i$ via the over-the-air
 computation \cite{Zhang&etal:arxiv2023}. As a result, the BS has the complex equivalent global model update $\tilde{\thetabf}_{m,t+1}$   given by
\begin{align}
\tilde{\thetabf}_{m,t+1} = \sum_{k\in\Kc_i}\rho_{k}\tilde{\thetabf}^{k,J}_{m,t} + \tilde{\nbf}^{\text{ul}}_{m,t} \label{eq_global_update_0}
\end{align}
where $\rho_k\in [0,1]$ is the weight with $\sum_{k\in\Kc_i}\rho_{k} = 1$, and $\tilde{\nbf}^{\text{ul}}_{m,t}\sim \mathcal{CN}({\bf 0}, \sigma^2_{\text{u}}\Ibf)$ is  the AWGN  at the BS receiver. The weight $\rho_k$ represents the uplink processing effect including device transmission and BS receiver processing.

For local model update  in \eqref{SGD}, let
$\Delta\tilde{\thetabf}^{k}_{m,t} \triangleq \tilde{\thetabf}^{k,J}_{m,t} - \tilde{\thetabf}^{k,0}_{m,t}$ denote the  equivalent complex representation of the
local model  difference after the local training at device $k\in\Kc_i$  in round $t$. Using  \eqref{dl_device_signal} and  \eqref{eq_global_update_0}, we can express the global model update $\tilde{\thetabf}_{m,t+1}$ from    $\tilde{\thetabf}_{m,t}$ as
\begin{align}
\!\!\!& \tilde{\thetabf}_{m,t+1}  = \tilde{\thetabf}_{m,t} + \sum_{k\in\Kc_i}\rho_{k}\Delta\tilde{\thetabf}^{k}_{m,t}
+ \sum_{k\in\Kc_i}\rho_{k}\tilde{\nbf}^{\text{dl}}_{k,t} +  \tilde{\nbf}^{\text{ul}}_{m,t} \nn\\
\!\!\!&\;\; + \sum_{j\neq i}\sum_{k\in\Kc_i}\!\rho_{k}\frac{\hbf^H_{k,n}\wbf^{\text{dl}}_{i,n}(\wbf^{\text{dl}}_{j,n})^H\hbf_{k,n}}{|\hbf^H_{k,n}\wbf^{\text{dl}}_{i,n}|^2}\!\cdot\!
\frac{\|\tilde{\thetabf}_{m,t}\|\bar{\thetabf}^{'}_{\hat{m}(j, t),t}}{\|\tilde{\thetabf}_{\hat{m}(j, t),t}\|}.
\label{eq_global_update}
\end{align}
Finally, the real-valued  global model update can be recovered from its complex version as
$\thetabf_{m,t+1}\!=[\Re{\{\tilde{\thetabf}_{m,t+1}\}^T}\!,$ $\Im{\{\tilde{\thetabf}_{m,t+1}\}^T}]^T$.

\section{Multi-Model Downlink  Beamforming Design}
\label{sec:upper_bound}

We consider the transmission design in the multi-model FL system to maximize the training convergence rate.
Recall that the BS transmits all $M$ model updates simultaneously to devices via multicast beamforming. Consider the global model update   $\thetabf_{m,nM}$ for model $m$ at the
beginning of each frame $n\in\Sc\triangleq\!\{0,\ldots,S-1\}$. We design the downlink beamforming vectors to  minimize
the maximum expected optimality gap to $\thetabf_m^\star$ among all $M$ models after $S$ frames, subject to the transmitter power budget. The optimization problem is formulated as
\begin{align}
\Pc_{o}: \,\, &\min_{\{\wbf^{\text{dl}}_{i,n}\}} \,\max_{m\in\Mc} \, \mae[\|\thetabf_{m,SM}- \thetabf_m^\star\|^2] \nn\\
&\text{s.t.}\,\,\,\,\, \sum^{M}_{i=1}\|\wbf^{\text{dl}}_{i,n}\|^2 \leq D_{\text{max}}P, \quad n\in \Sc \label{constra_dl_power}
\end{align}
where $\mae[\cdot]$ is  taken w.r.t.\ receiver noise and
mini-batch local data samples  at each device.
Problem $\Pc_{o}$ is a  stochastic optimization problem with a min-max objective. To tackle this challenging problem, we first develop a more tractable upper bound on $\mae[\|\thetabf_{m,SM}- \thetabf_m^\star\|^2]$
by analyzing the convergence rate of the global model update. Then, we propose a downlink multi-group multicast beamforming method to minimize this upper bound.
\begin{figure*}[t]
\centering
\hspace*{-0em}\includegraphics[scale=.56]{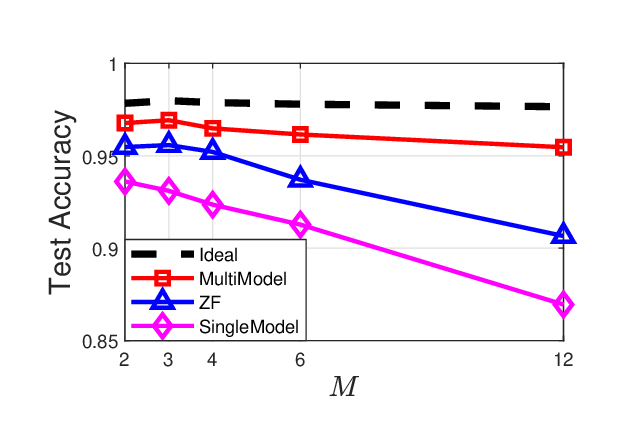}
\hspace*{-.8em}\includegraphics[scale=.56]{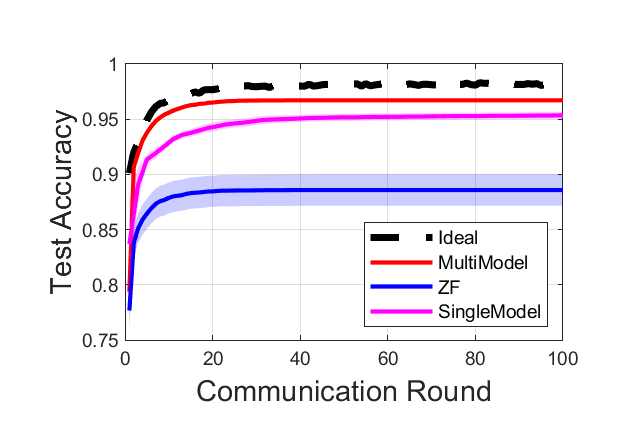}
\hspace*{-.8em}\includegraphics[scale=.56]{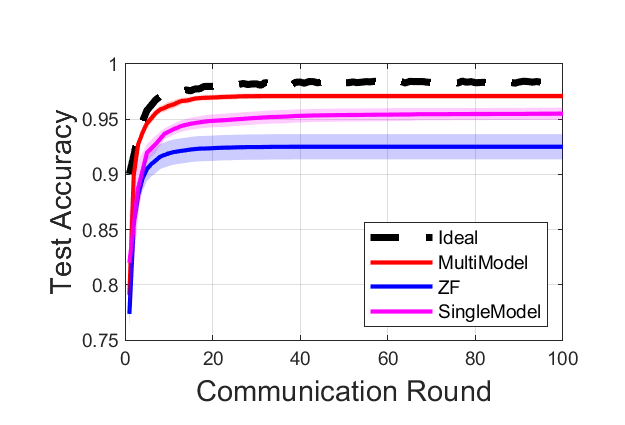}
\caption{Left: Test accuracy vs. $M$ (from Model A).
Middle \& Right: Test accuracy vs. communication round $t$: Middle -- Model A; Right -- Model B.}
\label{Fig:Accuracy_N128_K10_K12}
\end{figure*}
\subsection{Convergence Rate Analysis}

To analyze the model update convergence rate, we make the following  assumptions on
the local loss functions, the global model updates, and the difference between
the global and weighted sum of the local loss functions.
They are commonly assumed for the convergence
analysis of the FL model training \cite{Amiri&etal:TWC2022,Guo&etal:JSAC2022,Bhuyan&etal:2023}.

\begin{assumption}\label{assump_smooth}
The local loss function $F_m^{k}(\cdot)$ is $L$-smooth and $\lambda$-strongly convex,
$\forall m\in\Mc$, $\forall k\in\Kc_{\text{tot}}$.
\end{assumption}

\begin{assumption}\label{assump_bound_model}
Bounded model parameters:
$\|\tilde{\thetabf}_{m,t}\|^2 \leq r$,
for some $r>0$, $\forall m\in\Mc$, $\forall t$.
Bounded stochastic gradients and  sample-wise loss gradients:
$\mathbb{E}[\|\nabla F_m^{k}(\thetabf_m)\|^2] \leq \mu$,
$\|\nabla L_m(\thetabf_m;\sbf_{i},v_{i})\|^2 \leq \beta_{1}\|\nabla F_m^{k}(\thetabf_m)\|^2 + \beta_2$, for some $\mu>0$, $\beta_{1}\geq 1$
and $\beta_2 \geq 0$, $\forall m\in\Mc$, $\forall k\in\Kc_{\text{tot}}$, $\forall t,\forall i$.
\end{assumption}

\begin{assumption}\label{assump_bound_diff}
Bounded gradient divergence:
$\mae[\| \nabla F_m(\thetabf_{m,t}) - \sum_{k=1}^{K}c_{k}\nabla F^k_{m}(\thetabf^{k,\tau}_{m,t})  \|^2] \leq \phi$, for some $\phi \ge 0$, $0\leq c_k \leq 1$,
$\forall m\in\Mc$, $\forall \tau$, $\forall t$.
\end{assumption}

We now analyze the global model convergence rate over frames for each model $m$.
Based on  \eqref{eq_global_update}, we first obtain the per-model global update  over frames, \ie  $\thetabf_{m,nM}$.
Let device group   $\hat{i}$ be the group that trains model $m$ in communication round $t$ at frame $n. $ The device-model assignment  between $\hat{i}$ and $m$ is given in   \eqref{m(i)}. Summing both sides of \eqref{eq_global_update} over $M$ rounds in  frame $n$, and subtracting the complex version of the optimal $\tilde{\thetabf}_m^\star$ from both sides,
we obtain
\begin{align}
\tilde{\thetabf}_{m,(n+1)M} \!-\! \tilde{\thetabf}_m^\star \! = \!\tilde{\thetabf}_{m,nM}\! -\! \tilde{\thetabf}_m^\star\!
+\!\!\!\! \sum_{t=nM}^{(n+1)M-1}\!\!\!\!\!\sum_{k\in\Kc_{\hat{i}}}\!\!\rho_{k}\Delta\tilde{\thetabf}^{k}_{m,t}
\!+ \!\tilde{\ebf}_{m,n} \nn
\end{align}
where $\tilde{\ebf}_{m,n}$ is the accumulated error term  in \eqref{eq_global_update}  over $M$ rounds in frame $n$, given by
\begin{align}
& \tilde{\ebf}_{m,n} \triangleq  \sum_{t=nM}^{(n+1)M-1}\sum_{k\in\Kc_{\hat{i}}}\rho_{k}\tilde{\nbf}^{\text{dl}}_{k,t} +  \sum_{t=nM}^{(n+1)M-1}\tilde{\nbf}^{\text{ul}}_{m,t} \nn\\[-.5em]
& +\!\!\!\! \sum_{t=nM}^{(n+1)M-1}\!\!\!\!\sum_{k\in\Kc_{\hat{i}}}\rho_{k}\!\sum_{j\neq \hat{i}}\!\!\frac{\hbf^H_{k,n}\wbf^{\text{dl}}_{\hat{i},n}(\wbf^{\text{dl}}_{j,n})^H\hbf_{k,n}}{|\hbf^H_{k,n}\wbf^{\text{dl}}_{\hat{i},n}|^2}
\!\!\cdot\!\! \frac{\|\tilde{\thetabf}_{m,t}\|\bar{\thetabf}^{'}_{\hat{m}(j, t),t}}{\|\tilde{\thetabf}_{\hat{m}(j, t),t}\|}. \nn
\end{align}
By Assumption~\ref{assump_bound_model},
we can further bound $\mae[\|\tilde{\ebf}_{m,n}\|^2]$ by
\begin{align}
\mae\big[\|\tilde{\ebf}_{m,n}\|^2\big]\! \leq & \  rMK\sum_{i=1}^{M}\sum_{k\in\Kc_i}\frac{\sum_{j\neq i}|\hbf^H_{k,n}\wbf^{\text{dl}}_{j,n}|^2+ \tilde{\sigma}^2_{\text{d}}}{|\hbf^H_{k,n}\wbf^{\text{dl}}_{i,n}|^2}  \nn\\
& +M\tilde{\sigma}^2_{\text{u}} \nn
\end{align}
where $\tilde{\sigma}^2_{\text{d}} \triangleq \sigma^2_{\text{d}}D_{\text{max}}/2$, and $\tilde{\sigma}^2_{\text{u}} \triangleq \sigma^2_{\text{u}}D_{\text{max}}/2$.

Using the above, we  provide an upper bound on $\mae[\|\thetabf_{m,SM}- \thetabf_m^\star\|^2]$ in Proposition~\ref{thm:convergence} below. Due to the space limitation, the proof is omitted.
Part of the proof adopts some techniques in  \cite[Th. 2]{Bhuyan&etal:2023}.
\begin{proposition}\label{thm:convergence}
For the multi-model FL\ system described in Section~\ref{sec:FL_alg},  under  Assumptions~\ref{assump_smooth}--\ref{assump_bound_diff} and for  $\eta_n<\frac{1}{\lambda}$, $\forall n$, the expected model optimality gap after $S$ frames is upper bounded by
\begin{align}
\mae[\|\thetabf_{m,SM}- \thetabf_m^\star\|^2]\! \leq & \ \Gamma_m\!\prod_{n=0}^{S-1}\!G_n + \sum_{n=0}^{S-2}\!H(\wbf^{\text{dl}}_n)\!\!\!\prod_{s=n+1}^{S-1}\!\!G_s\nn\\
&   + H(\wbf^{\text{dl}}_{S-1}) +\Lambda, \quad m\in\Mc
\label{eq_thm1}
\end{align}
where $\Gamma_m \triangleq \mae[\| \thetabf_{m,0} - \thetabf_m^\star\|^2]$,  $G_n \triangleq 4(1-\eta_n\lambda)^{2J}$, $\Lambda \triangleq \sum_{n=0}^{S-2}C_{n} \big(\!\prod_{s=n+1}^{S-1}\!G_s\big)+ C_{S-1}$ with $C_{n} \!\triangleq 4\eta^2_{n}J^3K^2(\beta_1\mu+\beta_2)+4\eta^2_nJ^2\phi+4M\tilde{\sigma}^2_{\text{u}} $,
$\wbf^{\text{dl}}_n \triangleq [(\wbf^{\text{dl}}_{1,n}\!)^H, \ldots, (\wbf^{\text{dl}}_{M,n}\!)^H]^H$, and
\begin{align}
H(\wbf^{\text{dl}}_n) \triangleq 4rMK\sum_{i=1}^{M}\sum_{k\in\Kc_i}\frac{\sum_{j\neq i}|\hbf^H_{k,n}\wbf^{\text{dl}}_{j,n}|^2+ \tilde{\sigma}^2_{\text{d}}}{|\hbf^H_{k,n}\wbf^{\text{dl}}_{i,n}|^2}. \nn
\end{align}
\end{proposition}

\subsection{Downlink Multi-Group Multicast Beamforming Design}
\label{sec:dl_bf}
We now replace the objective function in $\Pc_{o}$
with  the more tractable upper bound in \eqref{eq_thm1}.
Note that only $\Gamma_m\!\prod_{n=0}^{S-1}\!G_n$
in \eqref{eq_thm1} depends on $m$. Omitting the constant terms $\Gamma_m\!\prod_{n=0}^{S-1}\!G_n +\Lambda$, we arrive at the following equivalent optimization problem w.r.t.\ multicast beamforming vectors  $\{\wbf^{\text{dl}}_n\}$ over $S$ frames:
\begin{align}
\Pc_{1}: \,\, &\min_{\{\wbf^{\text{dl}}_n\}^{S-1}_{n=0}} \,\, \sum_{n=0}^{S-2}\!H(\wbf^{\text{dl}}_n)\!\!\!\prod_{s=n+1}^{S-1}\!\!G_s  + H(\wbf^{\text{dl}}_{S-1}) & \text{s.t.} \quad \eqref{constra_dl_power}.\nn
\end{align}
Note that by Proposition~\ref{thm:convergence},  $G_n > 0$, for $\eta_n<\frac{1}{\lambda}$, $n\in\Sc$, and  $\prod_{s=n+1}^{S-1} G_s > 0$. Thus,  $\Pc_{1}$ can be decomposed into $S$ subproblems to solve,
one for each frame $n$, given by
\begin{align}
\Pc_{2,n}: \,\,&\min_{\wbf^{\text{dl}}_n} \,\, \sum_{i=1}^{M}\sum_{k\in\Kc_i}\frac{\sum_{j\neq i}|\hbf^H_{k,n}\wbf^{\text{dl}}_{j,n}|^2+ \tilde{\sigma}^2_{\text{d}}}{|\hbf^H_{k,n}\wbf^{\text{dl}}_{i,n}|^2} \nn\\
&\text{s.t.}\,\,\,\,\, \sum^{M}_{i=1}\|\wbf^{\text{dl}}_{i,n}\|^2 \leq D_{\text{max}}P. \nn
\end{align}

Problem $\Pc_{2,n}$ is a multi-group multicast beamforming problem with $M$ multicast beamforming vectors, one for each device group, to optimize. The objective is a total sum of interference-and-noise-to-signal ratios at the BS receiver as the result of downlink-uplink processing. The family of multicast beamforming problems is nonconvex and NP-hard \cite{Sidiropoulos&etal:TSP2006,Dong&Wang:TSP2020}.
We propose to use PGD to solve $\Pc_{2,n}$.    Since we can compute the  closed-form gradient updates fast,  PGD is suitable for solving $\Pc_{2,n}$.
Furthermore, it is  guaranteed to find an approximate stationary point of $\Pc_{2,n}$ in polynomial time \cite{Mokhtari&etal:NIPS2018,Zhang&etal:WCL2022}.

\section{Simulation Results}
\label{sec:simulations}
We consider the image classification task  under an LTE  system setting.
Following the typical LTE specifications,
we set the system  bandwidth  $10$~MHz and the maximum BS transmit power $47~\text{dBm}$.
The channels are generated i.i.d.\ as $\hbf_{k,t} = \sqrt{G_{k}}\bar{\hbf}_{k,t}$ with
$\bar{\hbf}_{k,t}\sim\mathcal{CN}({\bf{0}},{\bf{I}})$, and  $G_k$ being the path gain from the BS to device $k$,\ modeled as
$G_{k} [\text{dB}] = -169.2-35\log_{10}d_k - \psi_k$,
where the BS-device distance  $d_k\in(0.02, 0.5)$  km, and  $\psi_k$ represents  shadowing with standard deviation  $8~\text{dB}$.
Noise power spectral density is $N_0 = -174~\text{dBm/Hz}$,
with noise figure $N_F=8$~dB and $2$ dB at the device and  BS receivers, respectively.
We use the MNIST dataset \cite{Lecun&etal:MNIST} for the multi-model training and testing. It consists of $60,000$ training samples and $10,000$ test samples.
We consider training two types of convolutional neural networks:
i) \textbf{Model A}:  an $8\times3\times3$ ReLU convolutional layer,
a $2\times2$ max pooling layer, and a softmax output layer
with $13,610$ parameters.
ii) \textbf{Model B}:  an $8\times3\times3$ ReLU convolutional layer,
a $2\times2$ max pooling layer, a ReLU fully-connected layer with $20$ units, and a softmax output layer
with $27,350$ parameters.
The training samples are randomly and evenly distributed  across devices. The local dataset at device $k$ has $S_{k} = {60,000}/{K}$ samples.
For the local training using SGD at each device, we set  $\lambda=3$, $J=100$, mini-batch size $|\Bc^{k\tau}_{m,t}|={600}/{K}, \forall k,\tau,m,t$, and the learning rate $\eta_n=0.2$, $\forall n$.
We set weight $\rho_{k}=M/K, \forall k$ in the uplink model aggregation.
All results are obtained by
taking the current best test accuracy and averaging over $20$ channel realizations.

Besides our proposed method, denoted by MultiModel, we consider  three schemes for comparison:
i) \textbf{Ideal}: Perform multi-model FL\ via  \eqref{eq_global_update} with noise-interference-free downlink/uplink and perfect recovery of model parameters.
ii) \textbf{ZF}: Perform multi-model FL\ via  \eqref{eq_global_update} using the zero-forcing (ZF) multicast beamformers proposed in  \cite{Sadeghi&etal:2018TWC}.
iii) \textbf{SingleModel}: Use the single-model FL
 with downlink multicast beamforming for signal-to-noise-ratio maximization considered in (31) of  \cite{Zhang&etal:arxiv2023} to  train multiple models sequentially with $K$ devices.

Fig.~\ref{Fig:Accuracy_N128_K10_K12}-Left shows the test accuracy   after $30$ communication rounds vs.\ training  $M$ models, all from Model A. We set $(N,K) = (128,12)$. Our proposed MultiModel outperforms all other alternatives. Its performance remains roughly unchanged as $M$ increases and can achieve $\sim97\%$ test accuracy after $30$ rounds, while  other schemes noticeably deteriorate  as $M$ increases.
 Consider $M=2$, and one from Models A and B each. Fig.~\ref{Fig:Accuracy_N128_K10_K12}-Middle and Right show the test accuracy
over round $t$ for Models A and B, respectively. We set $(N,K) = (128,10)$.
The shadow area for each curve indicates the $90\%$ confidence interval.
MultiModel again outperforms other alternatives for both Models A and B.
Between  Models A and B, we see that Model B, which is the larger one, achieves
slightly higher test accuracy than Model A under both multi-model and single-model training.

\section{Conclusion}
Multi-model wireless FL with imperfect transmission/processing over noisy channels is considered in this paper. We formulate the downlink-uplink transmission process and obtain
the per-model global update expression in each round. We design downlink beamforming to maximize the FL training performance. Using an upper bound on the optimality gap of the global model update, we optimize downlink multicast beamforming for sending multiple models  simultaneously to device groups, which leads to a multi-group multicast beamforming problem for minimizing the sum of inverse received SINRs.
Simulation results demonstrate the effectiveness of the proposed multi-model method compared with the other alternatives.

\balance
\bibliographystyle{IEEEtran}
\bibliography{Refs}

\end{document}